**Improving essay peer grading accuracy in MOOCs using personalized weights from student's engagement and performance**

Carlos García-Martínez[1], Rebeca Cerezo[2], Manuel Bermúdez[3], Cristóbal Romero[1]

[1]University of Cordoba, Computer Science Department, Cordoba, Spain
[2]University of Oviedo, University of Oviedo, Psychology Department, Oviedo, Spain
[3]University of Cordoba, Social Science Department, Cordoba, Spain

Abstract

Most MOOC platforms either use simple schemes for aggregating peer grades, e.g., taking the mean or the median, or apply methodologies that increase students' workload considerably, such as *calibrated peer review*. To reduce the error between the instructor and students' aggregated scores in the simple schemes, without requiring demanding grading calibration phases, some proposals compute specific weights to compute a weighted aggregation of the peer grades. In this work, and in contrast to most previous studies, we analyse the use of students' engagement and performance measures to compute personalized weights and study the validity of the aggregated scores produced by these common functions, mean and median, together with two other from the information retrieval field, namely the geometric and harmonic means. To test this procedure we have analysed data from a MOOC about Philosophy. The course had 1059 students registered, and 91 participated in a peer review process that consisted in writing an essay and rating three of their peers using a rubric. We calculated and compared the aggregation scores obtained using weighted and non-weighted versions. Our results show that the validity of the aggregated scores and their correlation with the instructors grades can be improved in relation to peer grading, when using the median and weights are computed according to students' performance in chapter tests.

Keywords

# Introduction

Massiveness and low teaching involvement are two of the biggest problems for MOOCs. The huge enrolment in MOOCs can produce a student-teacher ratio of ten thousand to one or superior, so the time an instructor spends evaluating each student's work is very low or exceeding the capacity of a single instructor or teaching assistant (Chen, 2014). No single instructor could ever hope to grade assignments from so many students by him or herself. Several mechanisms have been proposed to assess assignments in MOOCs, including common Multiple-Choice Assessment (MCA), machine-based Automated Essay Scoring (AES) and Peer Review (PR) (Balfour, 2013). Peer Grading, Peer Assessment or Peer Review (PR) is a promising approach, with additional potential formative benefits in MOOCs (Suen, 2014). PR involves students in the evaluation process of correcting assignments. It uses a function to aggregate the different scores assigned by raters or peers. The median or arithmetic mean are commonly used in MOOCs and each one has strong and weak points, such as the mean uses all available data, while median is robust and it filters out extreme values. Others such as geometric mean and harmonic mean have also been used in similar contexts like Information Retrieval (Ravana and Moffat, 2009).

PR has also been criticized for being less rigorous than instructor assessment, too demanding on students and not reliable or fair due to student biases (Bachelet et al. 2015). In fact, reliability and validity of peer-generated grades are a major concern (Cho et al. 2008) not to mention their capability to calibrate their own preparedness (Alexander, 2013). *Reliability* refers to the consistency of peer scores, often measured by means of the inter-rater reliability. *Validity* is commonly calculated as the correlation coefficient between the aggregated peer scores and the instructor assigned ones, assuming that the instructor can provide fair and accurate grading marks. In order to calculate the correlation in validity investigations, the most common statistic is the Pearson's correlation. In this paper, we propose to improve the PR validity, i.e. the accuracy with regards to the instructor's grades, by exploiting information of the students' interaction with the MOOC's platform and their performance on chapter tests. However, peer raters do not participate equally for computing the qualification of another student. So, we propose to assign personalized weights according to the respective students' levels of engagement and performance in the MOOC. In this way, graders who show higher levels of engagement and obtain higher marks in the activities/quizzes of the course have associated higher weights than those who are not so engaged and obtain lower scores in the activities/quizzes of the course.

Although this work applies PR specifically in MOOC, it is in the general context of Computer-Assisted Learning/Instruction (CAL/CAI) that covers the whole range of uses of information and communication technology to support learning and knowledge exchange (Schittek et al., 2001). In fact, CAL/CAI has the goal of increasing the effectiveness and productivity teachers with the help of educational/computer technologies such as Peer Assessment or Peer Review (PR). The general related literature reports on many learning benefits (Albano et al. 2017) for peer-assessors like the exposure to different approaches, the development of self-learning abilities, the enhancement of critical thinking, etc., also reinforcing the development of student's explanation and argumentation processes. Moreover, even if it relies on grades assigned by intrinsically unreliable graders (the students), the application of peer grading also presents logistics advantages in saving teacher's time and providing fast feedback to the class. As we have mention previosly, one fundamental issue of peer grading is how to ensure that those grades are accurate, and how properly aggregate those peer grading to form a final consensus score for each students (Lynda et al., 2017). To address the above issue, this paper tries to improve the reliability of each grader by using personalized weights according to student's level of engagement and performance in the MOOC. In this way, we hope that our results can help to increase the use of PR not only in MOOC but also in others types of CAL/CAI environments. Based on this context, our study aim to provide answers to the next research questions:

**Question1:** May peer grading aggregation methods such as the geometric or harmonic mean provide more accurate scores with regards to those provided by the instructor?

**Question 2:** May the engagement and performance information, retrieved from the MOOC's platform, be useful to compute weights that improve the accuracy or validity of peer grading?

The paper is arranged in the following way. Firstly, the related background is described. Next, the used method is presented together with the description of the course and data. Then, the data analysis and discussion are described. Finally, conclusions and future work are dealt with.

## Literature overview

Peer grading, or peer assessment, is defined as "*an arrangement for learners to consider and specify the level, value, or quality of a product or performance of other equal-status learners*" (Bachelet et al. 2015). It is usually considered less rigorous and reliable than instructor assessment, apart from being too demanding on students who not only have to complete their assignments, but also evaluate some of their peers.

Many studies analyse how much accurate peer graders can be in relation with the instructor's grades. They usually report promising results with regard to the aggregation of the marks for low-order cognitive questions and not so promising ones if individual peer grades or higher-order questions were considered (Falchikov and Goldfinch, 2000; Kulkarni et al., 2014; Luo et al., 2014; Hafner & Hafner, 2003), even though students are almost always helped with scoring rubrics and incentived by making their final grades related to the quality of their peer evaluations.

In order to improve the accuracy of peer grades aggregations further, and thus lessen the need for many peer grades, several proposals exploit other information sources to compute specific weights for raters. The conceptual meaning of the weights can be expressed as (Torra, 2011): "*The weight is attached to the information source. Due to this, weights correspond to the importance of the information sources*". This implies that weights do not change the domain of the measure, the essay grading in our case, regardless how these weights were computed. And this is statistically consistent provided that weights have not got any unit of measure associated. Therefore, there is not any problem in using other student information to compute weights to aggregate the ratings, the output of the weighted aggregation of essay grading values is an essay grading estimation. But what are the most relevant pieces of information to scale the information sources, the raters. Table 1 categorizes the revised weights-based peer review approaches according to this secondary information used to obtain weights from, which has also been arranged according to the type of information (columns: quality of the evaluations, quality of the knowledge of the subject, and quality of the provided feedback) and how that was appraised (rows: with regard to the opinion of other peers, or with regard to the opinion of the instructor).

Next, we describe some works that consider weights in the aggregation function to get more accurate results. CrowdGrader (de Alfaro & Shavlovsky, 2014) assigns weights to graders that depend on the variance of their assigned scores, with regards to the scores received by the associated assignments. In PeerRank (Walsh, 2014), weights come from the grades that the own raters receive. This recursive definition requires the method to iterate several times until the aggregated scores converge to a stable state. Gutiérrez et al. (2014) construct a trust graph over the raters. Trust values, or weights, are firstly computed for raters that evaluated some assignments also evaluated by the instructor. Then, indirect trust values are computed for those who evaluated any assignment also evaluated by a rater with a trust value previously computed. Cho & MacArthuz (2010) and Cho et al. (2006) consider two stages. In the first stage, students submit and evaluate some preliminary assignments, providing a score and some feedback. Subsequently, students score the received feedback. The weights for a second stage are computed according to the scores of their feedback and differences between their grades and the aggregated score of the preliminary assignments. Piech et al (2013) present several probabilistic models (PG1, PG2, and PG3) that estimate the raters' biases, reliabilities and knowledge about the subject, from the provided scores, and correct the aggregated ones accordingly. *Calibrated peer review* (Pelaez, 2001; Russell, 2005) is another methodology, extensively applied, that also computes weights that improve the accuracy of PR. It introduces a calibration phase that helps students learn to grade by first practicing on example submissions, before evaluating real assignments of other students. Grade differences for example assignments, which the instructor had previously evaluated, are used to assign a weight to the students. The good results of calibrated peer review are counterbalanced by the fact that it increases the students' workload even more than the other methods.

Table 1 summarises the differences between previous approaches. For instance, Calibrated Peer Review considers the grade differences between raters and instructor's provided scores for practicing assignments, whereas the proposals in the first row do not require instructor's grades at any point. The other approaches either emphasize the knowledge that the raters have about the subject, according to the opinion of the other raters (first row, second column), or whether the provided feedback was positively evaluated by the rated student (first row, third column).

| **Whose is this information contrasted with?** | **What information is exploited?** | | |
|---|---|---|---|
| | How well raters evaluate | How much knowledge raters have about the subject | How useful raters' feedback was for the student |
| with other peers | • CrowdGrader (de Alfaro & Shavlovsky, 2014)<br>• Cho & MacArthur (2010)<br>• Piech et al, 2013<br>• Cho et al. (2006)<br>• Generalized PeerRank (Walsh, 2014) | • PeerRank (Walsh, 2014) | • Cho et al. (2006) |
| with the Instructor´s | • Gutiérrez et al. (2014)<br>• Calibrated Peer Review (Russell, 2005) | • Present approach | |

Table 1. Weights-based peer review approaches.

Finally, there is not any proposal, to our knowledge, that computes weights according to the knowledge raters have about the subject or whether the feedback was useful, contrasted with the instructor's opinion. The first case is covered by this present work, where the knowledge of the students is evaluated by means of repeated online multiple-choice tests. The second case is more complicated, since it would require the instructor to evaluate the feedback that raters provide to their peers, i.e., it magnifies the instructor's workload.

## Method

### Course description

We have used data from a Google Course Builder MOOC about Introduction to Philosophy (https://*omitted for double-blind review purposes*). The course was in Spanish and therefore, aimed at Spanish speaking students of all ages interested in Philosophy. To our knowledge, this is the first MOOC in the world about this topic in Spanish language. The purpose of the course was to introduce and disseminate basic philosophical content that may be of interest to the general public to help them to start thinking critically about life, politics, technology, religion, love, knowledge, etc. The course consisted of 7 chapters about the main/basic topics of philosophy. Each chapter had several Youtube video lectures as visual lessons, and one multiple-choice test with ten questions as assessment per chapter (value between 0 to 100). At the end of the whole course, students had to write an essay as final exam and to participate in a peer-review process in order to pass the course. The course started on April 2015 and lasted nine weeks: with one week per chapter plus two additional weeks, one week to write the essay and another week to do the peer review. The essay was about a topic explained during the course, which required students to introduce the topic, present different opinions, argue their own ones with examples and deductions, and provide a summarizing conclusion. So it is suitable to evaluate the critical thinking, analysis, and argumentation capabilities of the student who writes it. Essay shouldn't be longer than 1.000 words. All assignments were submitted online using the Google Course Builder platform. In it, as in other MOOC platforms, the peer review activity consists of two stages: submission and review. Students firstly submit an assignment to be peer reviewed and in turn, they must review three random assignments from peers. Students must complete the reviews in order to complete the course. Students were given a rubric that described the assessment criteria (see Appendix A).

Rubrics comprised guiding questions or dimensions according to which the student's work should be graded, and a 5-point Likert-scale quality gradation for each dimension, from poor (1-Poor) to excellent (5-Excellent). The sum of the four questions is the final grading score (a value from 0 to 20) that we have rescaled to a value in the interval from 0 to 10, which is the most typical scale in *omitted for double-blind review purposes*.

**Sample**

The course had 1059 students registered but only 91 completed the full course and participated in the peer-review process. According to self-reported demographic data, 53 of students were men (58.24%) and 38 were women (41.76%). The average age of students was 21.5. Given that the course was intended for Spanish speakers, it is natural that most of the students accessed from three Spanish-speaking countries: *omitted for double-blind review purposes* (44.61% students), *omitted for double-blind review purposes* (11.55%) and *omitted for double-blind review purposes* (10.59%).

The average number of chapters completed on time by the students was 5.5274 SD= 1.8698. Figure 1 shows the histogram or frequency of the number of chapters completed on time.

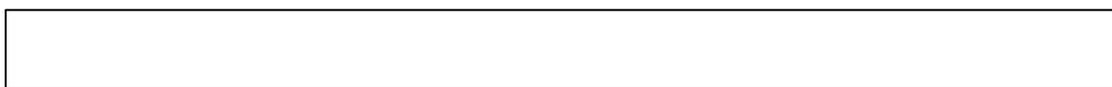

Figure 1: Histogram of number of completed chapters on time by students.

We can see in Figure 1 that there is a big group of highly engaged students (37.36%) who completed all the lessons of the course on time, each chapter during its week. It is followed by a big group of students (32.96%) who completed six lessons on time, and a small group of students (13.18%) who completed five lessons on time. On the other hand, there are three small groups of lowly engaged students (2.19%, 6.59% and 7.69%) who just completed three, two or one chapters on time, respectively.

The average mark of all students in the seven chapter tests of the course was 72.9449 out of 100 ±19.5253. Figure 2 shows the frequency of average marks obtained by the students.

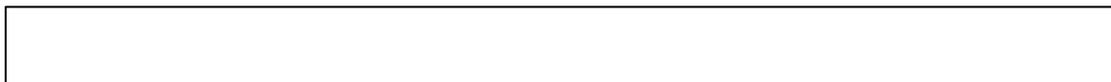

Figure 2: Histogram of students' mean marks in all the tests of the course.

We can see in Figure 2 that a great number of students obtained high marks in the tests. In fact, the most frequent averaged marks were in the range of 80 to 89 points (29.67%), followed by the range 70 to 79 points (18.68%), and the range 90 to 100 points (15.38%). The rest of the students (36.26%) obtained medium averaged marks in the range from 30 to 69 points.

## Data Analysis

### Procedure

The procedure that we used to collect the data is based on the rubric-aided essay peer-review. As aforementioned, the students had to do an essay as final exam and participate in the peer review process in order to pass the course. There were a total of 91 essays (N=91) that each received three peer grades. A few essays received less peer-grades (N=4) and were excluded from this study. Therefore, 95 students submitted their essays, but only 91 were reviewed by 3 raters (273 reviews) using the provided rubric (see Appendix 1). The instructor also graded the 91 assignments using the same rubric.

Our study assumes that the instructor provides the most accurate score for the assignments; therefore, the validity of peer-grading scores can be determined by their similarity to the instructor-assigned scores. We have used different aggregation methods to calculate a single score for each essay. We have used the following process to obtain up to twelve different aggregated scores, for each essay, in order to study and compare their validity with regard to the instructor grade (see Figure 3). Firstly, the instructor provided students with an evaluation rubric, which consists of five questions to be assessed according to a 5 Likert scale (it is described in the appendix A). Both the instructor and the students used this rubric to evaluate the submitted essay. To bound the students' workload, each assignment (essay) was assessed by a teacher and just three peer raters, so each student evaluated just three assignments. Then, different aggregation functions were applied on peers' grades to obtain a single measure to be compared with that of the teacher.

Eight aggregation functions have been considered. These functions are two common aggregation techniques widely applied in MOOCs (Bachelet et al. 2015), the mean and the median, and two from the field of information retrieval (Ravana and Moffat, 2009), the harmonic and geometric means, together with their weighted variants. Weights are computed according to the student's interaction and performance on the MOOC platform, as described at the end of this section.

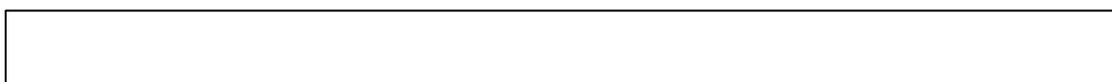

Figure 3: Process for obtained all the grades of a submitted essay.

Next, we formally define the used aggregation score methods. Given a set of $n$ observations S=$\{x_i \mid i=1,...,n\}$, the aggregation functions are:

- **Arithmetic mean or average:** It is the simplest and most commonly way to aggregate a set of observations of a phenomenon to obtain the central or typical tendency. On the other hand, it is not robust, given that it is greatly influenced by outliers. This means that the arithmetic mean may differ from one's notion of the typical tendency. The mean is computed as the sum of the observations divided by their cardinality:

$$\text{AM}(S) = \frac{1}{n} \sum x_i$$

- **Geometric mean:** The geometric mean is defined as the nth root of the product of the observations. It is always less or equal to the arithmetic mean, and more stable to outliers. On the other hand, if any of the observations is equal to zero, the geometric mean is also zero. In addition, it is advisable when averaging normalised results, i.e., it provides a single mean index from several heterogeneous sources with different ranges:

$$\text{GM}(S) = \sqrt[n]{\prod x_i}$$

- **Harmonic mean:** this aggregation measure is typically used for rates. It is defined as the reciprocal of the mean of the observations' reciprocals. It is always less or equal to the arithmetic and geometric means, given that it strongly tends toward the smallest observations, i.e., it mitigates the impact of large outliers and aggravates the impact of small ones:

$$\text{HM}(S) = \left( \frac{\sum x_i^{-1}}{n} \right)^{-1} = \frac{n}{\sum (1/x_i)}$$

- **Median:** The median of a set of observations is that which separates the higher half from the lower half of the observations, i.e., the value that occupies the central position when the observations are sorted. Its major benefits are that it is almost unaffected by outliers. For an even set of observations, it is defined as the average between the two middle observations:

$$\text{MD}(S_{\text{sorted}}) = \begin{cases} x_{(n+1)/2}, & \text{if n is odd} \\ \frac{\left( x_{n/2} + x_{n/2+1} \right)}{2}, & \text{if n is even} \end{cases}$$

Next, we describe the weighted extensions of previous measures, given the corresponding set of weights W={$w_i$ | i=1, …, n}:

- **Weighted (arithmetic) mean or average:** In contrast to the unweighted mean, this aggregation balances the contribution of each score according to its corresponding weight. Scores are multiplied by their weights and the sum is divided by the sum of the weights:

$$\text{Av(S,W)} = \frac{1}{\sum w_i} \sum w_i \, x_i$$

- **Weighted geometric mean:** In this case, weights are the powers of scores, and the exponent of the root is the sum of the weights:

$$\text{GM(S,W)} = \left( \prod x_i^{w_i} \right)^{1 \sum w_i}$$

- **Weighted harmonic mean:** this is defined as the reciprocal of the weighted average of the observations' reciprocals:

$$\text{HM(S,W)} = \left( \frac{\sum w_i \, x_i^{-1}}{\sum w_i} \right)^{-1} = \frac{\sum w_i}{\sum (w_i / x_i)}$$

- **Weighted median:** The weighted median is the 50% weighted percentile, i.e., the observation which occupies the central position when observations are sorted and each fills a space proportional to its weight (observations with weights equal to 0 are discarded):

$$\text{MD}(S_{\text{sorted}},\text{W}) = \left\{ \begin{array}{ll} x_t, & \text{if } \sum w_i < \frac{1}{2} \sum w_i < \sum w_i \\ \frac{x_t + x_{t+1}}{2}, & \text{if } \sum w_i = \frac{1}{2} \sum w_i \end{array} \right\}$$

Additionally, two sets of weights have been computed according to the engagement or the performance of the student in the MOOC platform:

- **Engagement-based weights**: They are proportional to the number of chapters students satisfactory completed on time during the course. Students had one week to complete each chapter before the next one starts. Then, the weight is the percentage of chapters that the student accessed, played its videos, and completed the associated quiz on time:

$$W^{\text{lessons}} = \frac{|\text{Lessons}_{\text{completed}}|}{|\text{Lessons}|}$$

- **Performance-based weights:** They are proportional to the marks obtained by students in all the chapter quizzes. It is calculated as the average mark (value between 0 to 100) obtained over the quizzes filled during the course:

$$W^{\text{quizzes}} = \frac{\sum \text{QuizScore}_i}{|\text{Quizzes}|}$$

## Results

Prior to the analysis, we visually inspect the score distributions for all the aggregation methods to assess their normality. This is done with the histograms (Figure 4) and boxplots (Figure 5) of the students' and instructor's grades.

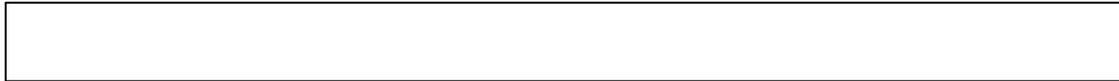

**Figure 4.** Frequency distributions and tendency of rubric-based scores for each traditional aggregation method versus that for the instructor's grades.

We can see in Figure 4 that there are not great differences between the distributions of the four aggregation methods, specially, between those of the Geometric and Harmonic means. It is interesting to see that the instructor provided more diverse marks than students. As we observe, there are more aggregated scores in the range from 6 to 9, whereas the instructor scored some of the corresponding essays out of that interval, either lower, from 4 to 6, or higher, from 9 to 10. In general, the mean of rubric-based instructor gradings is slightly lower, but with higher standard deviation (7.4340±1.306), than the rubric-based means of the four aggregation methods: Arithmetic mean (7.5753±0.9322), Geometric mean (7.5248±0.9557), Harmonic mean (7.4745±0.9821) and Median (7.5906±1.003). These results show that the scores given by the

MOOC students are, in general, a little higher than those given by the instructor and more biased towards a specific value around 7.5.

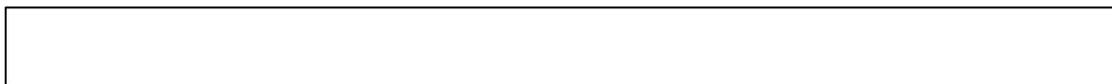

**Figure 5.** Boxplots of scores using all the aggregation methods versus instructor grading.

From Figure 5, we can see that all the distributions are very similar (especially those for the arithmetic, geometric and harmonic means) with median values around 7.5. However, engagement-based weighted aggregations are slightly higher. The instructor provides a variety of marks wider than those of unweighted and performance-based weighted aggregations. We think that this is a side effect of approximating a measure, the instructor's grades, by means of aggregating a biased random variable, the peers' grades. As known from the descriptive analysis, the students provided marks close to 7.5 very frequently. On the other hand, even though most of the median based aggregated scores condense in an interval narrower than those of the other aggregation functions, its maximal and minimal values are more distant than in the other cases. This is because the other metrics are more affected by distant values, concretely, students that have received two low marks and another much higher. In this case, the median, contrary to the other aggregation metrics, completely discards the highest mark.

Next, Pearson correlational analysis was used to analyse the similarity between the instructor's grades and the different proposed student's aggregation methods for grading, and among themselves as well. Table 2 shows the correlations among the traditional aggregation methods and the instructor's grades.

|  | Arithmetic mean | Geometric mean | Harmonic mean | Median | Instructor |
|---|---|---|---|---|---|
| **Arithmetic mean** | 1 | 0.9977*** | 0.9917*** | 0.9328*** | 0.5166*** |
| **Geometric mean** | 0.9977*** | 1 | 0.9980*** | 0.9280*** | 0.5228*** |
| **Harmonic mean** | 0.9917*** | 0.9980*** | 1 | 0.9175*** | 0.5266*** |
| **Median** | 0.9328*** | 0.9280*** | 0.9175*** | 1 | 0.4592*** |
| **Instructor** | 0.5166*** | 0.5228*** | 0.5266*** | 0.4592*** | 1 |

Table 2. Pearson's Correlation Coefficient between the four traditional aggregation methods in peer-grading and Instructor grades (***Correlations significant at 0.001 level 2-tailed).

On the one hand, Table 2 shows that there are a moderate positive correlation between the instructor grading and the four peer-grading aggregation scores. The three mean-based peer grading scores showed a little higher correlation (r=0.51 to r=0.52) than the median-based peer grading scores (0.45). On the other hand, we can see that the differences in correlation coefficient among the traditional aggregation methods are inconsequential, and the four types are highly correlated with each other (with r always above 0.91).

|  | Arithmetic mean | Geometric mean | Harmonic mean | Median |
|---|---|---|---|---|
| **Without weights** | 0.5166*** | 0.5228*** | 0.5266*** | 0.4592*** |
| **Engagement-based weights** | 0.6290*** | 0.6291*** | 0.6267*** | 0.6194*** |
| **Performance-based weights** | **0.7395*** | **0.7409*** | **0.7400*** | **0.7667*** |

Table 3. Pearson correlation of the 4 score aggregation methods, according to the weights considered, with regard to the instructor's scores (*** significance $p < .001$).

Table 3 shows the correlation between each aggregation method and the instructor's grades and how they change when weights are considered. We can see in Table 3, that the correlation increased one point (from about 0.5 to 0.6) in all the methods when we used the engagement-based weights and two points (from about 0.5 to 0.7) when we used the performance-based weights. So, correlation moves from moderate positive to strong positive. It is also interesting to see again that the Median was the method that improved the most by passing from the lowest correlation value, in the case of not using weights, to the highest correlation value when using the performance-based weights.

**Comparison with another proposal from the literature**

As previously addressed, *PeerRank* (Walsh, 2014) is a recent proposal for aggregating peer scores whose weighting scheme uses the knowledge about the subject that raters have, or in conjunction with the ability of raters to assign correct scores, *Generalized PeerRank* (see Table 1). As our proposal, PeerRank considers the knowledge of the raters to compute the weights from. The difference between PeerRank and the present proposal is that the raters´ knowledge is evaluated according to chapter tests, so according to the instructor's criteria, and PeerRank uses the own score aggregation that raters receive, i.e., their knowledge is evaluated according to their peers, and it being assumed that teachers are accurate in their assessments of students' level of success. This way, our experiment allows us to determine the precision improvement provided by the consideration of the instructor's opinion about that knowledge, in contrast with using the raters' opinions about that knowledge.

We computed the PeerRank and Generalised PeerRank aggregation schemes as indicated by Walsh (2014). Then, we compared the results of these schemes with the instructor' scores. Table 5 shows the correlation values for the PeerRank schemes and compares them to those of the arithmetic mean and median, with and without our performance-based weights (best results are boldfaced). Other weights and functions are not presented here for the sake of clarity.

| Aggregation approach | Pearson Correlation |
|---|---|
| PeerRank | 0.5129*** |
| Generalised PeerRank | 0.3806*** |
| Unweighted Arithmetic Mean | 0.5166*** |
| Unweighted Median | 0.4592*** |
| Perform.-based W. Arith Mean | 0.7395*** |
| Perform.-based W. Median | **0.7667***** |

Table 6. PeerRank versus some score aggregation approaches (*** significance $p < .001$).

We observe (see Table 6) that the results of PeerRank are comparable to those of the unweighted aggregations, and those of the performance-based weighted aggregations are much better. This reveals that evaluating the knowledge that raters have about the subject, used for aggregating the peers' scores, according to the instructor's criteria is more reliable than according to the peers' assigned scores. Additionally, we notice that the results of Generalised PeerRank are even worse than those of PeerRank. Given that their difference is that Generalised PeerRank also considers how accurate raters evaluated the other works, this suggests that this penalisation is just moving the final marks away from those provided by the instructor, who did not consider the accuracy of raters.

## Discussion

At this point, we are able to shed some light on the research questions; a series of different quantitative analysis was used to address:

**Question 1:** Can the use of other peer grading aggregation methods from other domains such as Information Retrieval provide more valid scores, i.e., more similar to those provided by the instructor? Regarding to this, we has not found significant differences between the new proposed aggregation methods from information retrieval: geometric and harmonic mean, and the two traditional ones in MOOCs, median and arithmetic mean. We have used boxplots and histogram to show that there are no significant differences among them. This result confirms that the traditionally used median and arithmetic mean are valid aggregation methods.

Regarding their accuracy, the median obtains the worst accuracy when no weights are used, but it gets the best accuracy when using performance-based weighted version. This is due to the fact that, even though the median is a very robust aggregation metric, the median of just three scores is extremely inaccurate if two of these scores differ from the real value. In this case, a mean aggregation represents better the three values, i.e., may get closer to the real value. However, once the significance of abnormal grades, provided by low performing students in our study, is reduced by the corresponding performance-based weights, the median provides a mark closer to real scores. In fact, it actually returns one of the provided scores, instead of an aggregation with several decimals. Similarly, Luo et al (2014) argued that the median is statistically more effective in dealing with extreme scores, which are expected to occur more often in the MOOC context, due to MOOC's students varying capabilities and/or motivation to grade.

**Question 2:** Can the use of weights improve the accuracy or validity of peer grading?

In the present study, the accuracy or validity has been improved when using engagement-based weights of 3 raters, and in a greater extent when using performance-based weights. Correlation was improved from 0.45 (moderate correlation) to 0.76 (strong correlation). Literature on peer grading offers contrasting views about accuracy in peer grading depending on the study and the number of used raters. Falchikov and Goldfich (2000) conducted a meta-analysis of 56 studies on peer grading in higher education and found a significant overall correlation (r=0.69) between student-assigned scores and instructor scores. In addition, multiple ratings were not found better than ratings by singletons. But other studies, such as Paré and Joordens (2008) obtained moderate correlations values (from 0.44 to 0.55) in two experimental studies to examine mark agreement between groups of experts and 5 peer markers using a Web-based assessment interface. However, a higher validity, very similar to the obtained in this work, it has been reported in most of the related MOOC studies about PR. Reilly et al. (2014) found a moderate correlation (0.57) between instructor and the average score of 3 raters. Bachelet et al. (2015) found an average correlation of 0.772 using 3-5 peers and an instructor. They also studied how many peer grades were required to correctly estimate best grade and they showed that a quick improvement was obtained going from 1 to 3 peer grades and the best appeared with 3-4 peer grades. Finally, Luo et al. (2014) found a significant correlation (r=0.619) between the instructor grading and the median-based peer grading scores when using 3 rates. They also slightly increased this correlation (r=0.662) when mean score, rather than median scores, was used to calculate the final grades.

Summarizing, in this work we have we have obtained a good correlation with instructor when using the performance-based weighted median for aggregating scores of 3 MOOC raters. We think that this high correlation can be due to the fact that in MOOCs only a few of students finish the course and carry out all the proposed actives such as quizzes, assessments, etc. (Romero & Ventura, 2017) and so, these students have both enough motivation and knowledge about the course to can be good raters.

## Conclusions

The present study is one of the first explorations in the context of open online education of how peer reviewer ability is related to students' performance assignment. We have analysed the validity of different aggregation methods, using and not using weights, for peer grading, which is based on a comparison between the aggregated students' scores and those assigned by an instructor. Our study was conducted using a sample consisting of 91 students who completed an Introduction to Philosophy MOOC and reviewed three essays using a rubric. So, a limitation of this study is that we always used 3 raters per assessment to limit the workload students had to carry out to pass the course. It is also important to notice that, compared with Calibrated Peer Review, our proposal reduces the student workload, because students do not need to undergo a calibration phase. The nature of the assessment task will influence validity of peer assessment. Assessments carried out in traditional academic areas within the classroom, like essays, have been observed to have better validities than those in areas of professional practice (e.g., intern performance, skills or practice) (Falchikov, & Goldfinch, 2000).

On the other hand, the main assumption underlying this study is that teachers are accurate in their assessments of students' level of success. At this moment some research suggests that assessment by multiple peers can even compete with assessment by an expert in terms of reliability (summative), feedback quality (formative), and subsequent improvement by the receiver, however there is not concluding results yet (Cho & MacArthur, 2010; Cho & Schunn, 2007; Cho, Schunn, & Charney, 2006). Therefore, PR validity was calculated according to the discrepancy between the aggregated scores and those of the instructor.

In sum, the results obtained in this work suggest two main implications about improving essay peer grading not only in MOOCs but also in other CAL/CAI environments. The first one is that the use of personalized student's weights from student's interaction information with MOOCs can improve peer grading accuracy. So, we recommend to use personalized/adaptive or configurable weights when aggregating student's peer grading scores as future direction in assessment approaches in MOOCs (Albano et al. 2017), (Alcarria et al., 2018). A promising research line is to automatically adjust those weights based on raters' past performance (Kulkarni et al. 2014) or by using a weight adjustment/optimization system (Villamañe et al. 2017). The second one is that performance-based weighted median scores can increase more the correlation with the instructor grade. So, we recommend obtaining different students' performance measures during the course in order to compute these weights. Although in this work we have assessed the student's performance by using chapter tests (traditional quizzes at the end of each chapter of the MOOC), it could be used other different form of assessment methods from online learning (Kearns, 2012) such as online discussions, fieldworks, presentations, etc. And these other assessment possibilities can open the door to future research about how to optimize the weights in peer grading. MOOC's provide much more information about the students, not only engagement in chapters and performance in test, such as information about the level of participation in forums, interaction with the videos, or the typically procrastinating variables in computer based learning environments (Cerezo, Sánchez-Santillán, Paule-Ruiz, & Núñez, 2016) that could be accurate resources to assess students performance. So, different types of personalized student's weights starting from different students' information can be studied for continuing improving accuracy of peer grading.

Finally, the degree to which our results can be generalized is limited. On one side, the available information on the MOOC participants in this study is limited; we have no information with respect to other participants' variables potentially related with this study (e. g. educational background, age, professional occupation, etc.). On the other side, only one MOOC was studied, therefore, future research on peer grading should include MOOCs with different course designs, topics and platforms. Despite the limitations, this study is going further to the fact that these finding may be of interest to developers of online peer assessment tools and open education platforms, given the increase in the number of MOOC, knowledge on how to optimize student learning experience with peer assessment is important for educational community involved in the use and design of MOOCs.

## Acknowledgements

### Conflict of Interest

Conflict of interest disclosure is not included according to submission instructions.

### Source of Funding

Source of Funding has been removed according to submission instructions. It should be included if the work was finally published.

Appendix

Appendix1: Grading Rubric

| | 5 (Excellent) | 4 (Very good) | 3 (Good) | 2 (Fair) | 1 (Poor) |
|---|---|---|---|---|---|
| **Writing** | Without mistakes in spelling, punctuation, and grammar | Very few mistakes in spelling, punctuation, and grammar | Few mistakes in spelling, punctuation, and grammar | Some mistakes in spelling, punctuation, and grammar | Spelling, punctuation or grammar errors seriously hinder comprehensibility |
| **Format and organization** | The logical structure of the argument is totally clear. Main arguments, objections and replies are very clearly indicated | The logical structure of the argument is very clear. Main arguments, objections and replies are very clearly indicated. | The logical structure of the argument is clear. Main arguments, objections and replies clearly indicated. | The structure of argument is not clear, or main arguments, objections and replies not clearly indicated | The structure of argument cannot be made out, or it is missing completely. |
| **Language and bibliographic** | It uses a language that is totally appropriate, without errors in citing materials, either in in-text citations or reference list | It uses a language that is quite appropriate, with minor errors in some references, either in in-text or reference list | It uses a language that is appropriate with few errors in references. | It uses a language not quite appropriate and it has got consistent errors either in in-text or reference list. | It uses a language clearly inappropriate and lacks understanding of proper citing procedures or lacks references altogether. |
| **Argumentation** | Concepts and argument clearly and accurately explained | Concepts and argument mostly clearly and accurately explained | Concepts and argument clear. | Concepts and argument not clearly and accurately explained | Concepts and argument not included or misrepresented |